# Structures of Cylindrical Ultrathin Copper Nanowires

Ho Jung Hwang and Jeong Won Kang*

Semiconductor Process and Device Laboratory, Department of Electronic Engineering, Chung-Ang University, 221 HukSuk-Dong, DongJak-Ku, Seoul 156-756, Korea

## ABSTRACT

To investigate cylindrical ultrathin copper nanowires, we performed atomistic simulations using the steepest descent method. The stable structures of the cylindrical ultrathin copper nanowires were multi-shell packs composed of coaxial cylindrical shells with {111}-like surfaces. Semiclassical orbits in a circle and circular rolling of a triangular network could explain the structures of the cylindrical ultrathin multi-shell copper nanowires. A calculation of the angular correlation function and the radial distribution function for nanowires showed that the structural properties of nanowires became closer to those of the bulk with increasing nanowire diameter.

# INTRODUCTION

Since the fundamental interest in low-dimensional physics and technological applications, such as for molecular electronic devices, has increased during the past decade, ultrathin metallic nanowires have been studied intensively [1-33]. In recent years, long metallic nanowires with well-defined structures having diameters of several nanometers have been fabricated using different methods [5-10]. Novel helical multi-shell structures have been observed in ultrathin gold nanowires [5-8], and these have been investigated using molecular dynamics (MD) simulations [11-16]. Multi-shell nanowires have also been made from several inorganic layered materials, such as $WS_2$, $MoS_2$, and $NiCl_2$ [17-19]. The cylindrical shells obtained in the MD simulation are similar to the geometric shells of clusters. The MD simulations have focused on infinite wires with periodic boundary conditions along the wire axes. For example, MD studies have been carried out on the structure of ultra-thin, infinite Pb and Al nanowires at T = 0 K [20, 21], on the pre-melting of infinite Pb nanowires orientated along the (110) direction [22], and on the melting of infinite Pt and Ag (100)-oriented nanowires [23]. The structures of freestanding Ti nanowires have been studied using a genetic algorithm and a tight-binding potential [24]. The strain rate effect induced by amorphizations of pure Ni and NiCu alloy nanowires [25] and the yielding and fracture mechanisms of Au and Cu nanowires [26, 27] have also been investigated using MD simulations. Yanson *et al.* have studied multi-shell structures in Na nanowires [28]. The stability of Na nanowires was studied by modeling them as infinite uniform jellium cylinders and by solving self-consistently [29]. In addition, the deformation and the breaking of atomic-sized sodium wires has been studied using a density functional simulation [30]. The stability of quasi-one-dimensional Si structures has been investigated using a generalized tight-binding MD scheme [31], and Si nanowires connected to Al electrodes have been studied using large-scale local density functional simulation [32]. Bilalbegović studied the room temperature structure of Al, Cu, and Au infinite nanowires using MD simulations and showed that cylindrical multi-shell and filled metallic nanowires exist for several fcc metals [16].

However, present knowledge of the structures and properties of metallic nanowires is still quite limited. In particular, as far as we know, there is no theoretical work on the atomic structures of cylindrical ultrathin copper nanowires. For cylindrical ultrathin copper nanowires, computer simulations can help in elucidating their properties and in developing new methods for their fabrication and can provide microscopic information on their physical properties. Therefore, in this research, we have investigated the structural properties of cylindrical ultathin copper nanowires.

## SIMULATION METHODS

The interaction between Cu atoms is described by a well-fitted potential function of the second moment approximation of the tight binding (SMA-TB) scheme [34]. The SMA-TB-type potential function has been used in atomistic simulation studies of nanoclusters [35-39] and ultrathin nanowires [24]. This potential is in good agreement with other potentials and with the experiments for bulk [34] and low-dimensional systems [33]. Table 1 shows the physical values for Cu calculated by using the SMA-TB scheme and other theological methods and measured by experiment. The physical values for Cu calculated by the SMA-TB scheme are in agreement with the other results.

In our study of the structure of cylindrical Cu nanowires, atomistic simulations were as follows: (1) We defined a cylinder with a diameter; then, an atom was inserted into the bottom of the cylinder. (2) Another atom was inserted into the bottom of the cylinder, and the atomic configuration was relaxed using the steepest descent (SD) scheme. (3) After sufficient relaxation, another atom was inserted into the bottom of the cylinder, and the atomic configuration was again relaxed using the SD scheme. This simulation was repeated until the length of the nanowire reached 40 Å. The reflective boundary condition (RBC) and the free boundary condition were then applied to the radial direction of the cylinders and to the axis of the nanowires, respectively. The diameter of the cylinder, $D_c$, ranged from 2 to 16 Å, and the positions of the atomic centers were sited along the radius, $D_c/2$.

## SIMULATION RESULTS AND DISCUSSION

Before discussion of our simulation results, we briefly review regular polygons inside a circle, which are applicable to cylindrical nanowires. A series of semiclassical orbits inside a circle is elementary geometry obtained by using the classical dynamics of a circular billiard ball on a circular billiard table [28, 40, 41]. Due to momentum conservation, only the ball's direction and position [40,41] determine its motion. The periodic orbits for the circular billiard ball are the regular polygons shown in Fig. 1. Each of these orbits can be characterized by a three-integer numbers $\beta(\nu, \omega, n)$, where $\nu$ is the number of turning points at the boundary during on period and $\omega$ measures how many times the trajectory encircles the center during the fundamental period. Therefore the winding number is $\omega$, and the number of vertices is $\nu$. Obviously, we have $\nu \geq 2\omega$. If there is a maximum common divisor $n$ between $\nu$ and $\omega$, the orbit is an $n$-fold repetition of a primitive periodic orbit (see (2,1,1), (4,2,2), (6,3,3), and (3,1,1), (6,2,2), and (9,3,3) in Fig. 1). Introducing an angle $\phi_{\nu\omega} = \pi \omega / \nu$, the length of a periodic orbit is $L_{\nu\omega} = 2 \nu R \sin\phi_{\nu\omega}$ from simple geometry, where $R$ is the radius. The structures of artificial cylindrical ultrathin Cu nanowires obtained from our simulations can be compared with the closed classical periodic orbits in a circular billiard with reflective walls.

Figure 2 shows some typical cylindrical Cu nanowire structures obtained from our simulations. In general, the stable structures of the cylindrical copper nanowires are multi-shell packs composed of coaxial cylindrical shells. The copper nanowires in some cases have a single-atom chain at their center. Each shell is formed by rows of atoms wound helically upwards, side by side. The pitch of the helices for the outer and the inner shells are different. The lateral surface of each shell exhibits a near-triangular network. Such helical multi-shell structures have been theoretically predicted for Al, Pb [20, 21], Au [11-16], and Ti nanowires [24] and recently experimentally observed in Au nanowires [5-8]. To characterize the multi-shell structures, Yanson *et al.* [28] used the classical orbits inside a circle and labeled the orbits of nanowires as (M,Q), where M is the number of vertices and Q is the winding number, and Kondo and Takayanagi [7] introduced the notation $n - n´ - n´´ - n´´´$ to describe a nanowire consisting of coaxial tubes with $n$, $n´$, $n´´$, $n´´´$ helical atom rows ($n > n´ > n´´ > n´´´$). Wang *et al* [24] used this notation. Since the thinnest magic nanowire consists of a single tube and a central strand, Tosatti *et al.* [13] used an ($n$, $h$) to denote a tube consisting of $n$ closely packed strands forming a maximal angle ranging from 30° ($n = 0$) to 0° ($h = n/2$) with respect to the tube axis. Although the index of Kondo and Takyanagi (*KT* index) is useful and easy to determine for multi-shell nanowire structures, the chirality information on nanowires cannot be characterized. The index of Tosatti *et al.* (*T* index) provides both chirality information and helical atom rows in the shells. In this work, we use both notations, the *KT* and the *T* indices, to denote cylindrical multi-shell nanowire structures.

The structures of the thinnest nanowire with *KT* index 2 ($D_c$ =2.56 Å), the nanowire with *KT* index 4 ($D_c$ = 2.8 Å - 3.2 Å), the 5 - 1 nanowire ($D_c$ = 4 Å), and the 6 - 1 nanowire ($D_c$ = 5 Å) in Fig. 2 are related to (2, 1, 1), (4, 1, 1), (5, 1, 1), and (6, 1, 1) of semiclassical orbits in Fig. 1, respectively. The 6 - 1, 8 - 3 ($D_c$ = 6 Å), 9 - 4 ($D_c$ = 7 Å), 11 - 6 - 1 ($D_c$ = 9 Å), 13 - 8 - 3 ($D_c$ = 10 Å), and 16 - 11 - 6 - 1 ($D_c$ = 12 Å) wires constitute a growth pattern with a five-atom difference between the shells. In contrast, the structure of the 20 - 16 - 10 - 5 - 1 ($D_c$ = 16 Å) wire constitute a growth pattern with a four-, five-, and six-atom difference between the shells, centered on a single atom chain. Helical multi-shell nanowires without a single atom chain in the center are 8 - 3, 9 - 4, and 13 - 8 - 3 wires. Each nanowire has a {111}-like surface. Previous simulation works on nanowire elongation deformation showed that a rectangular {100} nanowire transforms in to a cylindrical nanowire with a {111}-like surface by stretching [27]. In order to carry out a more detailed study of cylindrical multi-shell nanowire structures, we investigated the spreading sheets of nanowires in Fig. 2. Figure 3 shows the spreading sheets of nanowires, which are generally composed of a triangular network. To investigate the spreading sheets, we explain the *T* index of the triangular network sheet in Fig. 4. As shown in Fig. 4, the tube unit cell is given by the orthogonal vector (*n*, *h*) and the wire axis vector (*p*, *q*), in which *p* : *q* = (*n* - 2*h*) : (2*n* - *h*), and *h* ≤ *n*/2. All other tubes, except (*n*, 0) and (*n*, *h*/2), are chiral, and (*n*, *h*) and (*n*, *n*-*h*) are symmetrical with one another. At a constant *n* value, the wire center-to-center radius is given by $d_0$ $(n^2 + h^2 - nh)^{1/2} / 2\pi$ Å and decreases for increasing $h$, as the strands progressively align with the axis, where $d_0$ is the distance from the atomic center to the wire center. The total number of atoms per shell is $N = 2(n^2 + h^2 - nh)$, and the number in the central strand is $q$. In this paper, the central strand is denoted by (1, 1).

The nanowire structure indices of $KT$ and $T$ are shown in Table 2. Using the $KT$ index, we denote nanowires with $D_c$ = 2.8, 3.0, and 3.2 Å by 4. However, if the $T$ index is used, nanowires with $D_c$ = 2.8, 3.0, and 3.2 Å are denoted by (4, 2), (4, 1), and (4, 0), respectively. As mentioned above, at constant $n$ value, the wire diameter is linearly proportional to $d_0 (n^2 + h^2 - nh)^{1/2} / \pi$, and the chirality angle is given by $\tan^{-1}(\sqrt{3}h/(2n-1))$. In the case of $D_c$ = 16 Å, the spreading sheet of the first inner shell shows triangular and square networks, and the spreading sheet of the outer shell has one vacancy. Both semiclassical orbits in a circle and circular rolling of the triangular network can explain the structures of cylindrical ultrathin multi-shell copper nanowires.

We also analyze the angular correlation function (ACF) and the radial distribution function (RDF) relating to the structural properties of cylindrical ultrathin copper nanoiwres. Figure 5 shows the ACFs of cylindrical ultrathin copper nanowires. The dashed line, the case of $D_c$ = 16 Å, indicates the ACF of the bulk at 300 K. As the nanowire diameter is increased, the ACFs of the nanowires become similar to the ACF of the bulk. Since nanowires with 2.5 Å $\leq D_c \leq$ 3.5 Å have a rectangular structure, the peaks of the ACFs are different from the other ACFs at angle about 90°. In the 5 – 1 nanowire, main peaks at about 60° and 108° and a small peak at about 72° are observable, and the 108° and 72° peaks are related to the pentagonal structure. Since the shells of nanowires are composed of the triangular networks, the ACFs of most of the multi-shell structure nanowires have their main peaks at about 60°. Figure 6 shows the RDFs of cylindrical ultrathin Cu nanowires for different value of $D_c$. The dashed line in the case of $D_c$ = 16 Å indicates the RDF of the bulk at 300 K. As the nanowire diameter is increased, the RDFs of the nanowires become similar to the RDF of the bulk. In all nanowire cases, it is shown that the first nearest-neighbor atom distances are slightly closer than those in the bulk. Therefore, we also calculate the radii of cylindrical ultrathin Cu nanowires. Table III shows the radii of cylindrical ultrathin copper nanowires in Å. In Table III, A and B indicate nanowires obtained by our simulation and by rolling of a triangular network sheet, respectively. The nanowires obtained by rolling of a triangular network sheet are relaxed by using the steepest descent method. The values in brackets are radii calculated by using $d_0 \, (n^2 + h^2 - nh)^{1/2} / 2\pi$. The radius of each shell is the distance from the nanowire center. The values in the parentheses at the right of the *KT* index are the orthogonal vectors of the outer-shell for the nanowires. In the case of the 5-1 nanowire, the radius obtained from simulation and relaxation slightly increases more than the values in the brackets. However, in the other cases, the radii obtained from the simulation and relaxation are slightly less than those obtained by calculation. We think that is reason that, in the cross-sectional pentagon of the 5-1 nanowire, angles between the central strand and the outer-shell deviate from the angle of a normal triangle, as shown in Figs. 1 and 5.

**SUMMARY**

We investigated the structure of cylindrical ultrathin copper nanowires by using a cylinder and the steepest descent method. The stable structures of the cylindrical ultrathin copper nanowires are multi-shell packs composed of coaxial cylindrical shells. The theory of semiclassical orbits in a circle partially explained the properties of the structures of cylindrical ultrathin copper nanowires. An investigation of the spreading sheets of the nanowires obtained from our simulations showed that a coaxial cylindrical shell could be obtained by circular rolling of a triangular network sheet with an orthogonal vector. When the cylinder diameters are below 3.2 Å, the angular correlation functions show a main peak at about 90°. However, as the diameter of the nanowire is increased, the angular correlation and the radial distribution functions of the nanowires approach those of the bulk.

The structures of cylindrical ultrathin gold nanowires have been made clear by atomistic simulations and experiments; however, for the structures of cylindrical ultrathin copper nanowires, this work partially showed their properties by using atomistic simulations. Therefore, we are now preparing more specific theoretical and experimental works on such subjects as thermal effects, electronic properties, nanowire fabrication using STM and TEM images, thus overcoming our confined simulation work on copper nanowires.

**Table Caption**

Table I.  SMA-TB results are compared to results of a simple analytic nearest-neighbor embedded-atom model developed by Johnson [42] and Baskes [43]. The SMA-TB results for a surface diffusion barrier are compared to other results, such as effective medium (EM)[45, 46], corrected effective medium (CEM) [47], embedded-atom models in the Voter-Chen parameterization (EA(VC)) [48] and in the Adams-Foiles-Wolfer parameterization (EA(AFW)) [49], *ab-initio* density-functional calculations in the local-density approximation (LDA) [50], *ab-initio* density-functional results with gradient corrections (GGA) [50], and experimental results [51]. $E_c$ and $a_0$ are the cohesive energy of the atom and the lattice constant, respectively.

Table II. Structure index of the cylindrical ultrathin copper nanowires obtained by our simulations. $D_c$ is the diameter of the cylinder, the *KT* index is the index of Kondo and Takyanagi [7], and the *T* index is the index of Tosatti *et al.* [13].

Table III. Radii of the cylindrical ultrathin copper nanowires; A and B are for nanowires obtained by our simulation and for relaxed nanowires of structures obtained by rolling of triangular networks sheet, respectively. The values in brackets [ ] are the radii calculated by using $d_0 (n^2 + h^2 - nh)^{1/2} / 2\pi$, where $d_0$ is the atomic distance in the bulk at T = 0 K and (*n*, *h*) is the orthogonal vector of the nanowire. The unit of the radii is Å.

**Figure Captions**

Figure 1. Series of semiclassical inscribed inside a circular cross section, which are applicable to spheres (clusters) and cylinders (nanowires) alike.

Figure 2. Morphologies of cylindrical ultrathin Cu nanowires with diameters form 2 Å to 16 Å. In each case, a top view (left) and a side view (right) are presented.

Figure 3. Spreading sheets of each cylindrical Cu nanowire case in Fig. 2

Figure 4. Triangular network sheet. An ($n$, $h$) shell is made by cylindrically rolling a triangular lattice. A ($p$, $q$) is the axis vector of the ($n$, $h$) shell. $p : q = (n\text{-}2h) : (2n\text{-}h)$, and $h \leq n/2$.

Figure 5. Angular correlation functions for nanowires.

Figure 6. Radial distribution functions for nanowires.

# TABLES

**Table I.**

|  | $E_c$(eV) | $a_0$(Å) | Energy/atom on surface (eV) | | | Energy/atom of nanowire (eV) | | |
| --- | --- | --- | --- | --- | --- | --- | --- | --- |
|  |  |  | (100) | (110) | (111) | Surface | edge | inside |
| **SMA-TB** | -3.544 | 3.615 | -2.999 | -2.913 | -3.127 | -2.999 | -2.547 | -3.548 |
| **EAM-I** | -3.540 | 3.620 | -2.989 | -2.848 | -3.064 | -2.989 | -2.499 | -3.451 |
| **EAM-II** | -3.540 | 3.620 | -2.360 | -2.065 | -2.655 | -2.360 | -1.475 | -3.540 |
|  | SMA-TB | EM | CEM | EA(VC) | EA(AFW) | LDA | GGA | Experiment |
| **Diffusion Barriers on (100) surface (eV)** | | | | | | | | |
| **Jump** | 0.41 | 0.40[b] | 0.47[d] | 0.53[f] | 0.38[f] | 0.65-0.75[g] | 0.51-0.55[g] | 0.39±0.06[h] |
| **Exchange** | 0.79 | - | 0.43[d] | 0.79[f] | 0.72[f] | 1.03-1.23[g] | 0.82-0.96[g] | - |
| **Diffusion Barriers on (110) surface (eV)** | | | | | | | | |
| **In channel** | 0.23[a] | 0.29[c] | 0.26[e] | 0.53[f] | 0.24[f] | - | - | - |
| **Cross channel** | 0.29[a] | 0.56[c] | 0.49[e] | 0.31[f] | 0.30[f] | - | - | - |

[a] Ref. [44].
[b] Ref. [45].
[c] Ref. [46].
[d] Ref. [47].
[e] Ref. [48].
[f] Ref. [49].
[g] Ref. [50].
[h] Ref. [51].

Table II.

| $D_c$ (Å) | Structure indexes | |
|---|---|---|
| | *KT* index | *T* index |
| | n - n' - n'' - n'''-n'''' | orthogonal vectors |
| 2.5 | 2 | (2,0) |
| 2.8 | 4 | (4,2) |
| 3.0 | 4 | (4,1) |
| 3.2 | 4 | (4,0) |
| 4.0 | 5 - 1 | (5,0)(1,1) |
| 5.0 | 6 - 1 | (6,0)(1,1) |
| 6.0 | 8 - 3 | (8,1)(3,1) |
| 7.0 | 9 - 4 | (9,1)(4,2) |
| 9.0 | 11 - 6 - 1 | (11,0)(6,0)(1,1) |
| 10.0 | 13 - 8 - 3 | (13,1)(8,1)(3,1) |
| 12.0 | 16 - 11 - 6 - 1 | (16,0)(11,0)(6,0)(1,1) |
| 16.0 | 20 - 16 - 10 - 5 - 1 | (20,0)(16,0)(10,0)(5,0)(1,1) |

Table III.

| 5-1 (5,0) | | 6-1 (6,0) | | 16-11-6-1 (16,0) | |
|---|---|---|---|---|---|
| **A** | **B** | **A** | **B** | **A** | **B** |
| [0 - 2.037] | | [0 - 2.445] | | [0 - 2.445 - 4.482 - 6.519] | |
| 0 - 2.09 | 0 - 2.10 | 0 - 2.40 | 0 - 2.42 | 0 - 2.36 - 4.40 - 6.47 | 0 - 2.36 - 4.39 - 6.46 |

**FIGURES**

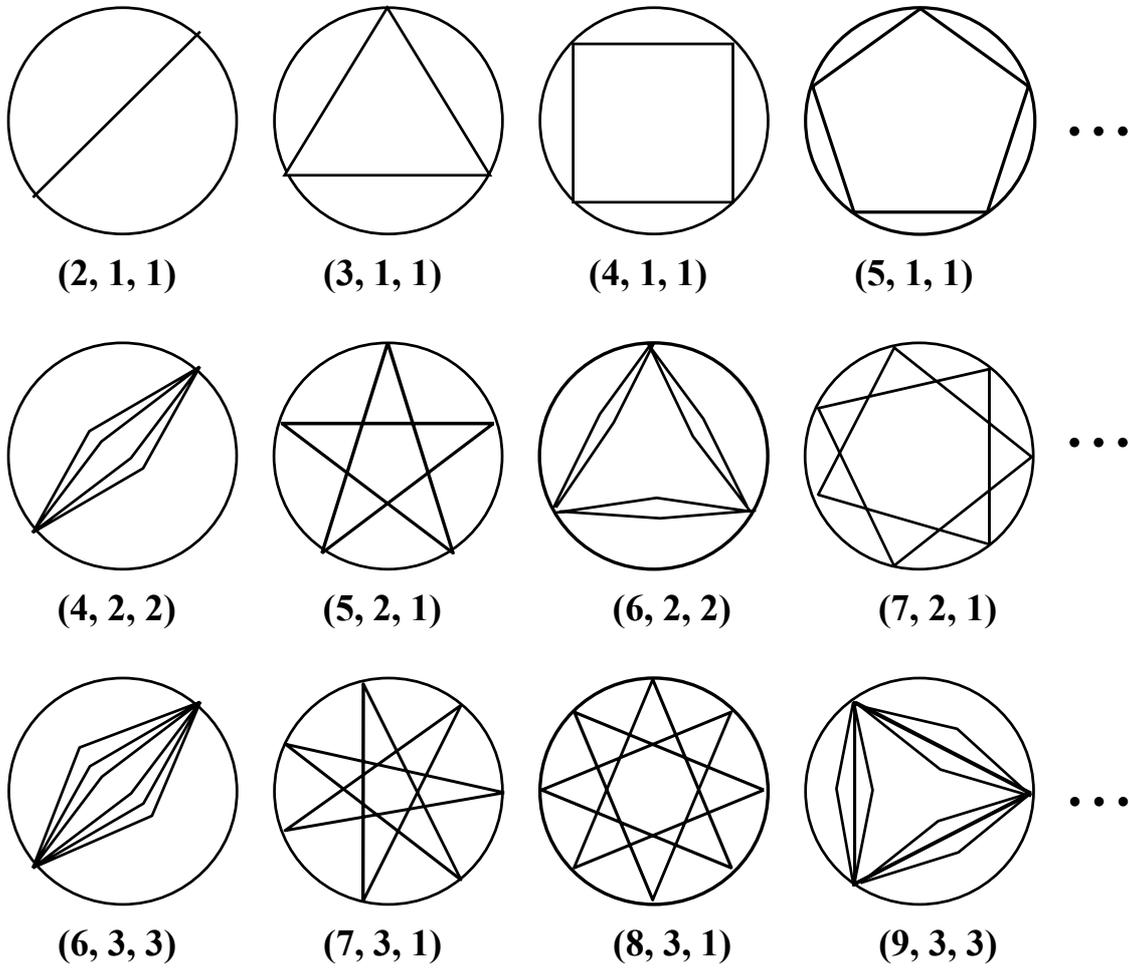

Figure 1.

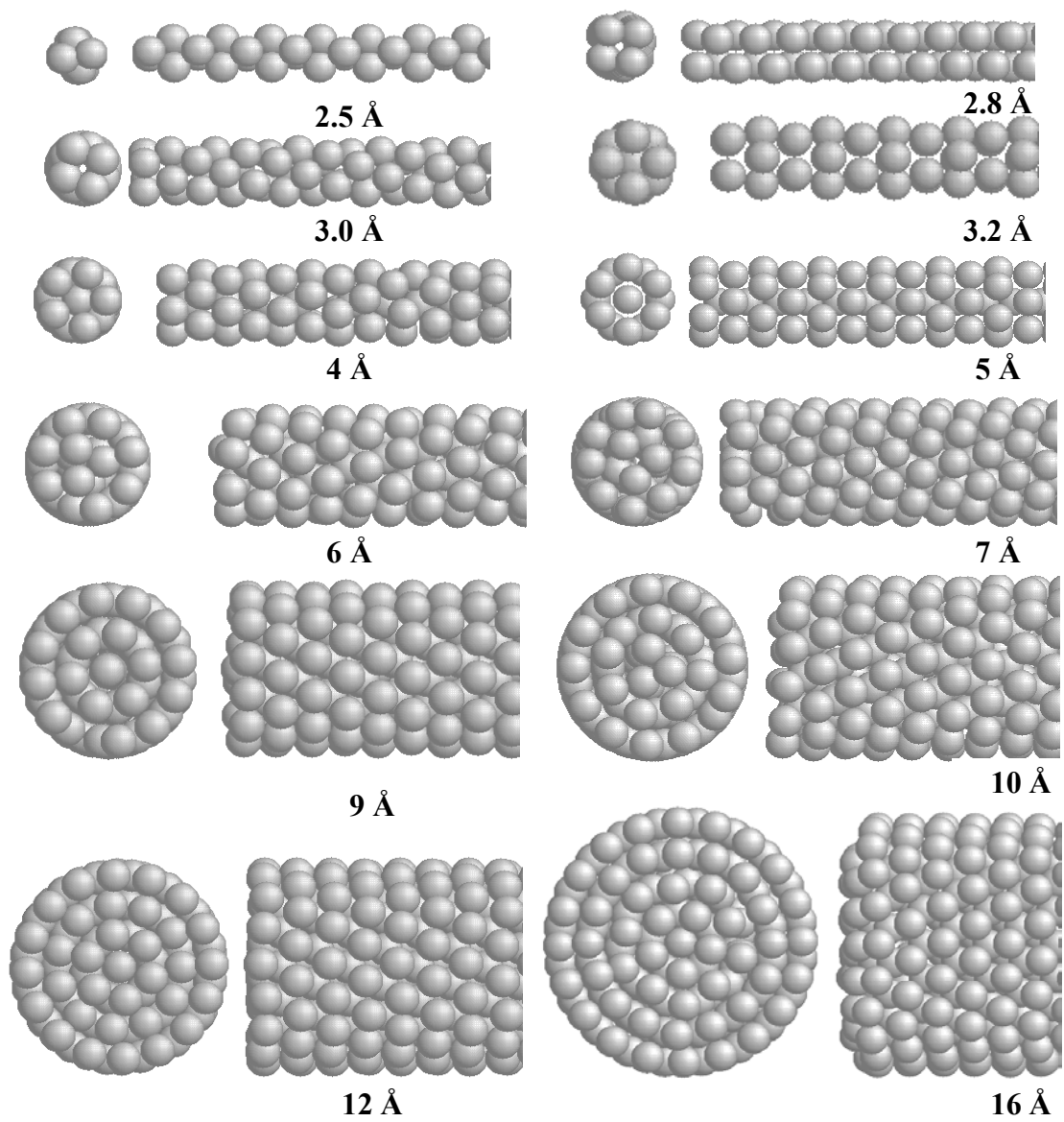

Figure 2.

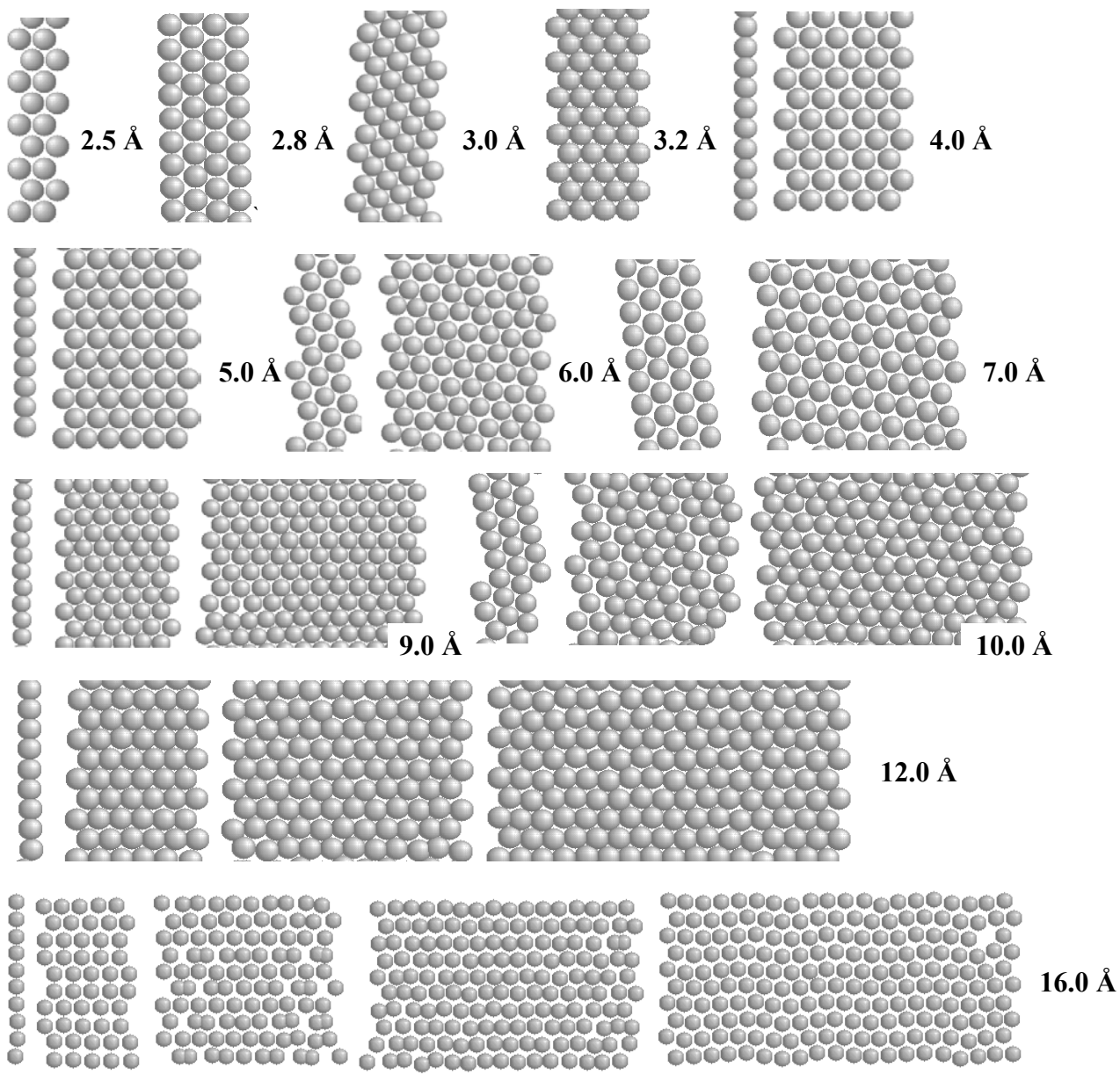

Figure 3.

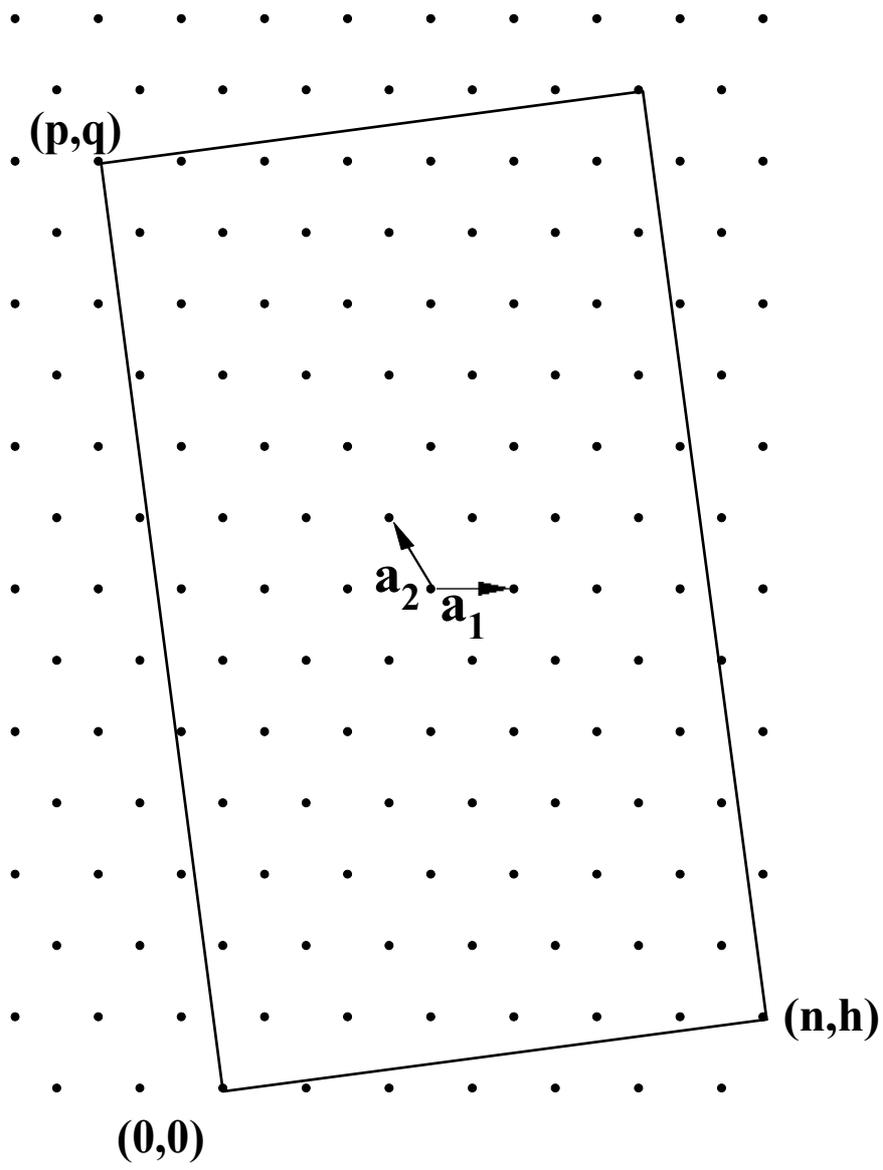

Figure 4.

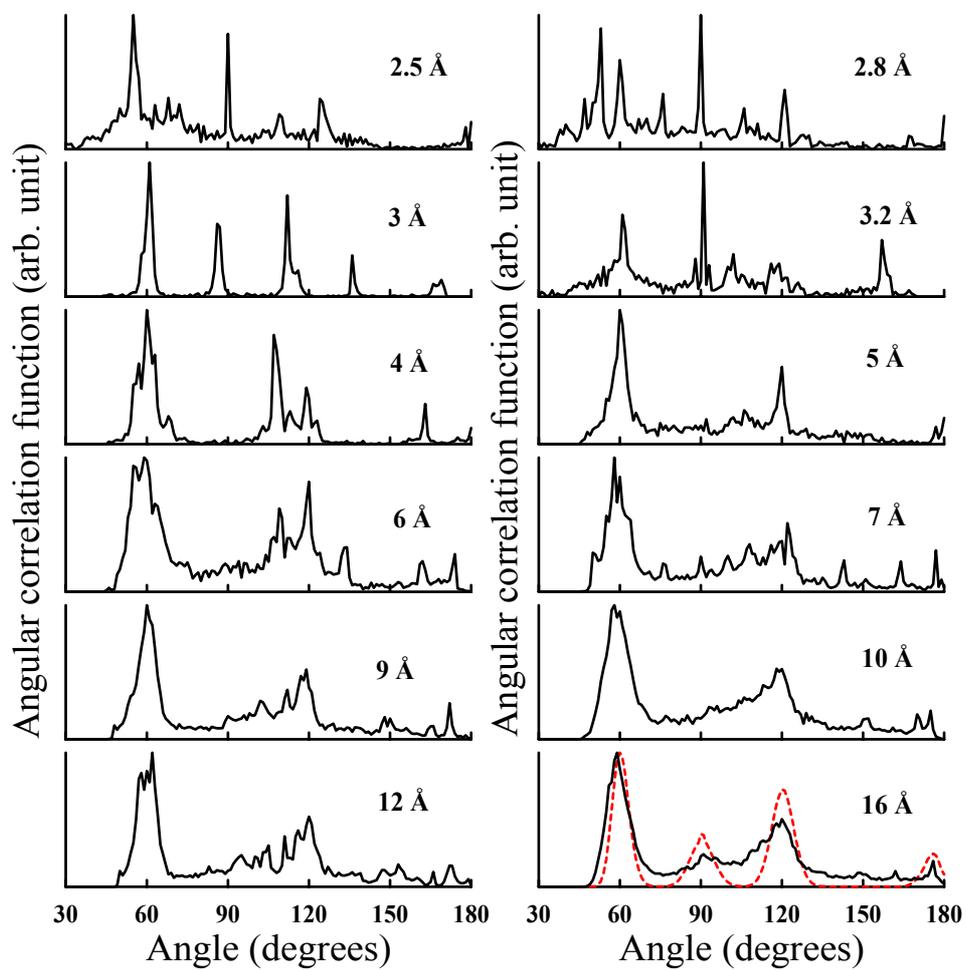

Figure 5.

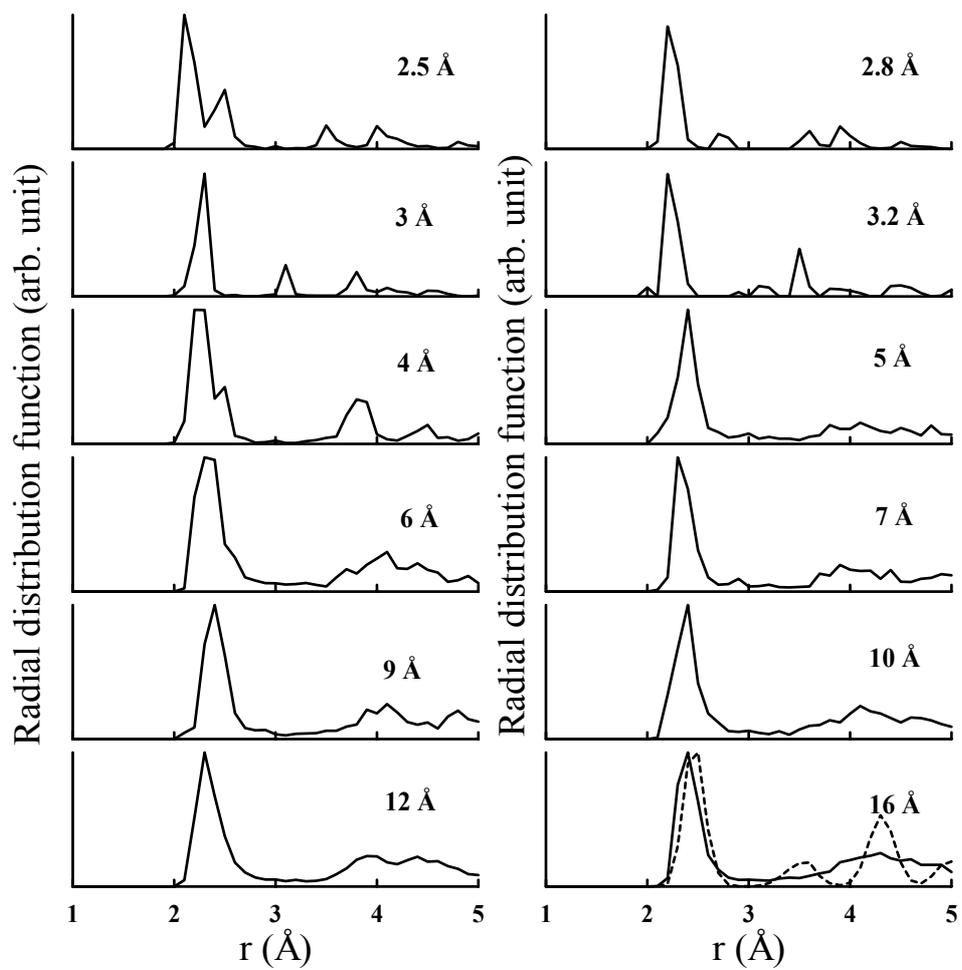

Figure 6.